\documentclass[11pt]{revtex4} 
\usepackage{amssymb,epsf}
\usepackage{latexsym}
\begin{document}
\title{Interacting holographic dark energy in Brans-Dicke theory}
\author{Ahmad Sheykhi \footnote{sheykhi@mail.uk.ac.ir}}
\address{Department of Physics, Shahid Bahonar University, P.O. Box 76175, Kerman, Iran\\
         Research Institute for Astronomy and Astrophysics of Maragha (RIAAM), Maragha,
         Iran}
 \begin{abstract}
We study cosmological application of interacting holographic
energy density in the framework of Brans-Dicke cosmology. We
obtain the equation of state and the deceleration parameter of the
holographic dark energy in a non-flat universe. As system's IR
cutoff we choose the radius of the event horizon measured on the
sphere of the horizon, defined as $L=ar(t)$. We find that the
combination of Brans-Dicke field and holographic dark energy can
accommodate $w_D = -1 $ crossing for the equation of state of
\textit{noninteracting} holographic dark energy. When an
interaction between dark energy and dark matter is taken into
account, the transition of $w_D$ to phantom regime can be more
easily accounted for than when resort to the Einstein field
equations is made.

\end{abstract}
\maketitle

\section{Introduction\label{Int}}
Recent data from type Ia supernova, cosmic microwave background
(CMB) radiation, and other cosmological observations suggest that
our universe is currently experiencing a phase of accelerated
expansion and nearly three quarters of the universe consists of
dark energy with negative pressure \cite{Rie}. Nevertheless, the
nature of such a dark energy is still the source of much debate.
Despite the theoretical difficulties in understanding dark energy,
independent observational evidence for its existence is
impressively robust. Explanations have been sought within a wide
range of physical phenomena, including a cosmological constant,
exotic fields, a new form of the gravitational equation, new
geometric structures of spacetime, etc, see \cite{Pad} for a
recent review. One of the dramatic candidate for dark energy, that
arose a lot of enthusiasm recently, is the so-called ``Holographic
Dark Energy" (HDE) proposal. This model is based on the
holographic principle which states that the number of degrees of
freedom of a physical system should scale with its bounding area
rather than with its volume \cite{Suss1} and it should be
constrained by an infrared cutoff \cite{Coh}. On these basis, Li
\cite{Li} suggested the following constraint on its energy density
$\rho_D \leq3c^2m^2_p/L^2$, the equality sign holding only when
the holographic bound is saturated. In this expression $c^2$ is a
dimensionless constant, $L$ denotes the IR cutoff radius and
$m^2_p =(8\pi G)^{-1}$ stands for the reduced Plank mass. Based on
cosmological state of holographic principle, proposed by Fischler
and Susskind \cite{Suss2}, the HDE models have been proposed and
studied widely in the literature
\cite{Huang,Hsu,HDE,Setare,Seta1,Setare1}. The HDE model has also
been tested and constrained by various astronomical observations
\cite{Xin,Feng} as well as by the anthropic principle
\cite{Huang1}. It is fair to claim that simplicity and
reasonability of HDE model provides more reliable frame to
investigate the problem of dark energy rather than other models
proposed in the literature. For example, the coincidence problem
can be easily solved in some models of HDE based on the
fundamental assumption that matter and HDE do not conserve
separately \cite{Pav1}.

On the other side, scalar-tensor theories of gravity have been
widely applied in cosmology \cite{Fara}. Scalar-tensor theories
are not new and have a long history. The pioneering study on
scalar-tensor theories was done by Brans and Dicke several decades
ago who sought to incorporate Mach's principle into gravity
\cite{BD}. In recent years this theory got a new impetus as it
arises naturally as the low energy limit of many theories of
quantum gravity such as superstring theory or Kaluza-Klein theory.
Because the holographic energy density belongs to a dynamical
cosmological constant, we need a dynamical frame to accommodate it
instead of general relativity. Therefore it is worthwhile to
investigate the HDE model in the framework of the Brans-Dicke
theory. The studies on the HDE model in the framework of
Brans-Dicke cosmology have been carried out in
\cite{Pavon2,Setare2,other}. The purpose of the present paper is
to construct a cosmological model of late acceleration based on
the Brans-Dicke theory of gravity and on the assumption that the
pressureless dark matter and HDE do not conserve separately but
interact with each other. Given the unknown nature of both dark
matter and dark energy there is nothing in principle against their
mutual interaction and it seems very special that these two major
components in the universe are entirely independent. Indeed, this
possibility is receiving growing attention in the literature
\cite{Ame,Zim,wang1,wang2,wang3} and appears to be compatible with
SNIa and CMB data \cite{Oli}. On the other hand, although it is
believed that our universe is spatially flat, a contribution to
the Friedmann equation from spatial curvature is still possible if
the number of e-foldings is not very large \cite{Huang}. Besides,
some experimental data has implied that our universe is not a
perfectly flat universe and recent papers have favored the
universe with spatial curvature \cite{spe}.

In the light of all mentioned above, it becomes obvious that the
investigation on the interacting HED in the framework of non-flat
Brans-Dicke cosmology is well motivated. We will show that the
equation of state of dark energy can accommodate $w_D = -1 $
crossing. As systems's IR cutoff we shall choose the radius of the
event horizon measured on the sphere of the horizon, defined as
$L=ar(t)$. Our work differs from that of Ref. \cite{Pavon2} in
that we take $L=ar(t)$ as the IR cutoff not the Hubble radius
$L=H^{-1}$. It also differs from that of Ref. \cite{Setare2}, in
that we assume the pressureless dark matter and HDE do not
conserve separately but interact with each other, while the author
of \cite{Setare2} assumes that the dark components do not interact
with each other.

This paper is outlined as follows: In section \ref{HDE}, we
consider noninteracting HDE model in the framework of Brans-Dicke
cosmology in a non-flat universe. In section \ref{INTHDE}, we
extend our study to the case where there is an interaction term
between dark energy and dark matter. We summarize our results in
section \ref{CONC}.

\section{HDE in Branse-Dicke cosmology\label{HDE}}
The action of Brans-Dicke theory is given by
\begin{equation}
 S=\int{
d^{4}x\sqrt{g}\left(-\varphi {R}+\frac{\omega}{\varphi}g^{\mu
\nu}\partial_{\mu}\varphi \partial_{\nu}\varphi +L_M
\right)}.\label{act0}
\end{equation}
The above action can be transformed into the standard canonical
form by re-defining the scalar field $\varphi$ and introducing a
new field $\phi$, in such a way that
\begin{equation}
 \varphi=\frac{\phi^2}{8\omega}.
\end{equation}
Therefore, in the canonical form, the action of Brans-Dicke theory
can be written \cite{Arik}
\begin{equation}
 S=\int{
d^{4}x\sqrt{g}\left(-\frac{1}{8\omega}\phi ^2
{R}+\frac{1}{2}g^{\mu \nu}\partial_{\mu}\phi \partial_{\nu}\phi
+L_M \right)},\label{act1}
\end{equation}
where ${R}$ is the scalar curvature and $\phi$ is the Brans-Dicke
scalar field. The non-minimal coupling term $\phi^2 R$  replaces
with the Einstein-Hilbert term ${R}/{G}$ in such a way that
$G^{-1}_{\mathrm{eff}}={2\pi \phi^2}/{\omega}$,  where
$G_{\mathrm{eff}}$ is the effective gravitational constant as long
as the dynamical scalar field $\phi$ varies slowly. The signs of
the non-minimal coupling term and the kinetic energy term are
properly adopted to $(+---)$ metric signature. The HDE model will
be accommodated in the non-flat Friedmann-Robertson-Walker (FRW)
universe which is described by the line element
\begin{eqnarray}
 ds^2=dt^2-a^2(t)\left(\frac{dr^2}{1-kr^2}+r^2d\Omega^2\right),\label{metric}
 \end{eqnarray}
where $a(t)$ is the scale factor, and $k$ is the curvature
parameter with $k = -1, 0, 1$ corresponding to open, flat, and
closed universes, respectively. A closed universe with a small
positive curvature ($\Omega_k\simeq0.01$) is compatible with
observations \cite{spe}. Varying action (\ref{act1}) with respect
to metric (\ref{metric}) for a universe filled with dust and HDE
yields the following field equations
\begin{eqnarray}
 &&\frac{3}{4\omega}\phi^2\left(H^2+\frac{k}{a^2}\right)-\frac{1}{2}\dot{\phi} ^2+\frac{3}{2\omega}H
 \dot{\phi}\phi=\rho_M+\rho_D,\label{FE1}\\
 &&\frac{-1}{4\omega}\phi^2\left(2\frac{{\ddot{a}}}{a}+H^2+\frac{k}{a^2}\right)-\frac{1}{\omega}H \dot{\phi}\phi -\frac{1}{2\omega}
 \ddot{\phi}\phi-\frac{1}{2}\left(1+\frac{1}{\omega}\right)\dot{\phi}^2=p_D,\label{FE2}\\
 &&\ddot{\phi}+3H
 \dot{\phi}-\frac{3}{2\omega}\left(\frac{{\ddot{a}}}{a}+H^2+\frac{k}{a^2}\right)\phi=0,
 \label{FE3}
\end{eqnarray}
where $H=\dot{a}/a$ is the Hubble parameter, $\rho_D$ and $p_D$
are, respectively, the energy density and pressure of dark energy.
We further assume the energy density of pressureless matter can be
separated as $\rho_M=\rho_{BM}+\rho_{DM}$,  where $\rho_{BM}$ and
$\rho_{DM}$ are the energy density of baryonic and dark matter,
respectively. We also assume the holographic energy density has
the following form
\begin{equation}\label{rho1}
\rho_{D}= \frac{3c^2\phi^2 }{4\omega L^2},
\end{equation}
where $\phi^2={\omega}/({2\pi G_{\mathrm{eff}}})$. In the limit of
Einstein gravity where $G_{\mathrm{eff}}\rightarrow G$, the above
expression reduces to the holographic energy density in standard
cosmology
\begin{equation}\label{rhoEins}
\rho_{D}= \frac{3c^2}{8\pi G L^2}=\frac{3c^2m^2_p}{L^2}.
\end{equation}
The radius $L$ is defined as
\begin{equation}\label{L}
L=ar(t),
\end{equation}
where the function $r(t)$ can be obtained from the following
relation
\begin{equation}
 \int_{0}^{r(t)}{\frac{dr}{\sqrt{1-kr^2}}}=\int_{0}^{\infty}{\frac{dt}{a}}=\frac{R_h}{a}.
\end{equation}
It is important to note that in the non-flat universe the
characteristic length which plays the role of the IR-cutoff is the
radius $L$ of the event horizon measured on the sphere of the
horizon and not the radial size $R_h$ of the horizon. Solving the
above equation for the general case of the non-flat FRW universe,
we have
\begin{equation}
r(t)=\frac{1}{\sqrt{k}}\sin y,\label{rt}
\end{equation}
where $y=\sqrt{k} R_h/a$. Now we define the critical energy
density, $\rho_{\mathrm{cr}}$, and the energy density of the
curvature, $\rho_k$, as
\begin{eqnarray}\label{rhocr}
\rho_{\mathrm{cr}}=\frac{3\phi^2 H^2}{4\omega},\hspace{0.8cm}
\rho_k=\frac{3k\phi^2}{4\omega a^2}.
\end{eqnarray}
We also introduce, as usual, the fractional energy densities such
as
\begin{eqnarray}
\Omega_M&=&\frac{\rho_M}{\rho_{\mathrm{cr}}}=\frac{4\omega\rho_M}{3\phi^2
H^2}, \label{Omegam} \\
\Omega_k&=&\frac{\rho_k}{\rho_{\mathrm{cr}}}=\frac{k}{H^2 a^2},\label{Omegak} \\
\Omega_D&=&\frac{\rho_D}{\rho_{\mathrm{cr}}}=\frac{c^2}{H^2L^2}.
\label{OmegaD}
\end{eqnarray}
For latter convenience we rewrite Eq. (\ref{OmegaD}) in the form
\begin{eqnarray}
HL=\frac{c}{\sqrt{\Omega_D}}. \label{HL}
\end{eqnarray}
Taking derivative with respect to the cosmic time $t$ from Eq.
(\ref{L}) and using Eqs. (\ref{rt}) and (\ref{HL}) we obtain
\begin{eqnarray}
\dot{L}=HL+a\dot{r}(t)=\frac{c}{\sqrt{\Omega_D}}-\cos y.
\label{Ldot}
\end{eqnarray}
Consider the FRW universe filled with dark energy and pressureless
matter which evolves according to their conservation laws
\begin{eqnarray}
&&\dot{\rho}_D+3H\rho_D(1+w_D)=0,\label{consq}\\
&&\dot{\rho}_M+3H\rho_M=0, \label{consm}
\end{eqnarray}
where $w_D$ is the equation of state parameter of dark energy. We
shall assume that Brans-Dicke field can be described as a power
law of the scale factor, $\phi\propto a^\alpha$. A case of
particular interest is that when $\alpha$ is small whereas
$\omega$ is high so that the product $\alpha \omega$ results of
order unity \cite{Pavon2}. This is interesting because local
astronomical experiments set a very high lower bound on $\omega$
\cite{Will}; in particular, the Cassini experiment implies that
$\omega>10^4$ \cite{Bert,Aca}. Taking the derivative with respect
to time of relation $\phi\propto a^\alpha$ we get
\begin{eqnarray}\label{dotphi}
&&\dot{\phi}=\alpha H\phi, \\
&&\ddot{\phi}=\alpha^2H^2\phi+\alpha\phi\dot{H}.\label{ddotphi}
\end{eqnarray}
Taking the derivative  of Eq. (\ref{rho1}) with respect to time
and using Eqs. (\ref{Ldot}) and (\ref{dotphi}) we reach
\begin{eqnarray}
\dot{\rho}_D=2H\rho_D\left(\alpha-1+\frac{\sqrt{\Omega_D}}{c}\cos
y\right)\label{rhodot}.
\end{eqnarray}
Inserting this equation in conservation law (\ref{consq}), we
obtain the equation of state parameter
\begin{eqnarray}
w_D=-\frac{1}{3}-\frac{2\alpha}{3}-\frac{2\sqrt{\Omega_D}}{3c}\cos
y\label{wD}.
\end{eqnarray}
It is important to note that in the limiting case  $\alpha=0$
($\omega\rightarrow\infty$), the Brans-Dicke scalar field becomes
trivial and Eq. (\ref{wD}) reduces to its respective expression in
non-flat standard cosmology \cite{Huang}
\begin{eqnarray}
w_D=-\frac{1}{3}-\frac{2\sqrt{\Omega_D}}{3c}\cos y\label{wDstand}.
\end{eqnarray}
We will see that the combination of the Brans-Dicke field and HDE
brings rich physics. For $\alpha\geq 0$, $w_D$ is bounded from
below by
\begin{eqnarray}
w_D=-\frac{1}{3}-\frac{2\alpha}{3}-\frac{2\sqrt{\Omega_D}}{3c}\label{wDbound}.
\end{eqnarray}
If we take $\Omega_D= 0.73$  for the present time and choosing
$c=1$ \cite{c}, the lower bound becomes $
w_D=-\frac{2\alpha}{3}-0.9$. Thus for $\alpha=0.15$ we have
$w_D=-1$.  The cases with $\alpha
>0.15$ and $\alpha <0.15$
 should be considered separately. In the first case where $\alpha
>0.15$ we have  $w_D<-1$. This is an interesting result and shows that, theoretically, the combination of
Brans-Dicke scalar field and HDE can accommodate $w_D = -1 $
crossing for the equation of state of dark energy. Therefore one
can generate phantom-like equation of state from a
\textit{noninteracting} HDE model in the Brans-Dicke cosmology
framework. This is in contrast to the general relativity where the
equation of state of a noninteracting HDE cannot cross the phantom
divide \cite{Li}. In the second case where $0\leq\alpha<0.15$ we
have $-1< w_D\leq-0.9$. Since $\alpha \approx 1/{\omega}$ and for
$\omega\geq 10^4$ the Brans-Dicke theory is consistent with solar
system observations \cite{Bert}, thus practically $\alpha\simeq
10^{-4}$ is compatible with recent cosmological observations which
implies $w_D\simeq -0.903$ for the present time in this model. In
both cases discussed above $w_D<-1/3$ and the universe undergoing
a phase of accelerated expansion. It is worthwhile to note that
since $\alpha\approx1/\omega$ and $\omega> 10^4$, therefore for
all practical purposes Brans-Dicke theory reduces to Einstein
gravity as one can see from the above discussion.

For completeness, we give the deceleration parameter
\begin{eqnarray}
q=-\frac{\ddot{a}}{aH^2}=-1-\frac{\dot{H}}{H^2},
\end{eqnarray}
which combined with the Hubble parameter and the dimensionless
density parameters form a set of useful parameters for the
description of the astrophysical observations. Dividing  Eq.
(\ref{FE2}) by $H^2$, and using Eqs. (\ref{rho1}), (\ref{HL}),
(\ref{dotphi}) and (\ref{ddotphi}), we find
\begin{eqnarray}
q=\frac{1}{2\alpha+2}\left[(2\alpha+1)^2+2\alpha(\alpha\omega-1)+\Omega_k+3\Omega_D
w_D\right]\label{q1}.
\end{eqnarray}
Substituting $w_D$ from Eq. (\ref{wD}), we get
\begin{eqnarray}
q=\frac{1}{2\alpha+2}\left[(2\alpha+1)^2+2\alpha(\alpha\omega-1)+\Omega_k-(2\alpha+1)\Omega_D-\frac{2}{c}{\Omega^{3/2}_D}\cos
y \right]\label{q2}.
\end{eqnarray}
If we take $\Omega_D= 0.73$ and $\Omega_k\approx 0.01$  for the
present time and choosing $c=1$, $\alpha\omega\approx1$,
$\omega=10^4$ and $\cos y\simeq 1$, we obtain $q=-0.48$ for the
present value of the deceleration parameter which is in good
agreement with recent observational results \cite{Daly}. When
$\alpha \rightarrow0$, Eq. (\ref{q2}) restores the deceleration
parameter for HDE model in Einstein gravity \cite{wang2}
\begin{eqnarray}
q=\frac{1}{2}(1+\Omega_k)-\frac{\Omega_D}{2}-\frac{\Omega^{3/2}_D}{c}\cos
y\label{q3}.
\end{eqnarray}
\section{Interacting HDE in Branse-Dicke cosmology\label{INTHDE}}
In this section we extend our study to the case where both dark
components- the pressureless dark matter and the HDE- do not
conserve separately but interact with each other. Although at this
point the interaction may look purely phenomenological but
different Lagrangians have been proposed in support of it
\cite{Tsu}. Besides, in the absence of a symmetry that forbids the
interaction there is nothing, in principle, against it. Further,
the interacting dark energy has been investigated at one quantum
loop with the result that the coupling leaves the dark energy
potential stable if the former is of exponential type but it
renders it unstable otherwise \cite{Dor}. Therefore, microphysics
seems to allow enough room for the coupling. With the interaction
between the two dark constituents of the universe, we explore the
evolution of the universe. The total energy density satisfies a
conservation law
\begin{equation}\label{cons}
\dot{\rho}+3H(\rho+p)=0.
\end{equation}
where $\rho=\rho_{M}+\rho_{D}$ and $p=p_D$. However, since we
consider the interaction between dark energy and dark matter,
$\rho_{DM}$ and $\rho_{D}$ do not conserve separately. They must
rather enter the energy balances \cite{Pav1}
\begin{eqnarray}
&&
\dot{\rho}_D+3H\rho_D(1+w_D)=-Q,\label{consq2}\\
&&\dot{\rho}_{DM}+3H\rho_{DM}=Q, \label{consm2}\\
&&\dot{\rho}_{BM}+3H\rho_{BM}=0,\label{consbm}
\end{eqnarray}
where we have assumed the baryonic matter does not interact with
dark energy. Here $Q$ denotes the interaction term and we take it
as $Q =3b^2 H(\rho_{DM}+\rho_{D})$ with $b^2$ is a coupling
constant. This expression for the interaction term was first
introduced in the study of the suitable coupling between a
quintessence scalar field and a pressureless cold dark matter
field \cite{Ame,Zim}. The choice of the interaction between both
components was meant to get a scaling solution to the coincidence
problem such that the universe approaches a stationary stage in
which the ratio of dark energy and dark matter becomes a constant.
In the context of HDE models, this form of interaction was derived
from the choice of Hubble scale as the IR cutoff \cite{Pav1}.

Combining Eqs. (\ref{rhocr}) and (\ref{dotphi}) with the first
Friedmann equation (\ref{FE1}), we can rewrite this equation as
\begin{eqnarray}\label{rhos}
\rho_{\mathrm{cr}}+\rho_k=\rho_{BM}+\rho_{DM}+\rho_D+\rho_{\phi},
\end{eqnarray}
where we have defined
\begin{eqnarray}\label{rhophi}
\rho_{\phi}\equiv\frac{1}{2}\alpha
H^2\phi^2\left(\alpha-\frac{3}{\omega}\right).
\end{eqnarray}
Dividing Eq. (\ref{rhos}) by $\rho_{\mathrm{cr}}$, this equation
can be written as
\begin{eqnarray}\label{Fried2new}
\Omega_{BM}+\Omega_{DM}+\Omega_D+\Omega_{\phi}=1+\Omega_k,
\end{eqnarray}
where
\begin{eqnarray}\label{Omegaphi}
\Omega_{\phi}=\frac{\rho_{\phi}}{\rho_{\mathrm{cr}}}=-2\alpha
\left(1-\frac{\alpha \omega}{3}\right).
\end{eqnarray}
Thus, we can rewrite the interaction term $Q$ as
\begin{eqnarray}\label{Q}
Q=3b^2H(\rho_{DM}+\rho_D)=3b^2H\rho_D(1+r),
\end{eqnarray}
where $r={\rho_{DM}}/{\rho_D}$ is the ratio of the energy
densities of two dark components,
\begin{eqnarray}\label{r}
r=\frac{\Omega_{DM}}{\Omega_D}=-1+\frac{1}{\Omega_D}\left[1+\Omega_k-\Omega_{BM}
+2\alpha\left(1-\frac{\alpha\omega}{3}\right)\right].
\end{eqnarray}
Using the continuity equation (\ref{consbm}), it is easy to show
that
\begin{eqnarray}\label{m}
\Omega_{BM}=\Omega_{BM0}a^{-3}=\Omega_{BM0}(1+z)^{3},
\end{eqnarray}
where $\Omega_{BM0}\approx0.04$ is the present value of the
fractional energy density of the baryonic matter and $z=a^{-1}-1$
is the red shift parameter. Inserting Eqs. (\ref{rhodot}),
(\ref{Q}) and (\ref{r}) in Eq. (\ref{consq2}) we obtain the
equation of state parameter
\begin{eqnarray}
w_D=-\frac{1}{3}-\frac{2\alpha}{3}-\frac{2\sqrt{\Omega_D}}{3c}\cos
y-b^2
{\Omega^{-1}_D}\left[1+\Omega_k-\Omega_{BM}+2\alpha\left(1-\frac{\alpha
\omega}{3}\right)\right]\label{wDInt}.
\end{eqnarray}
If we define, following \cite{Setare1}, the effective equation of
state as
\begin{eqnarray}\label{wef}
w^{\mathrm{eff}}_D=w_D+\frac{\Gamma}{3H},
\end{eqnarray}
where $\Gamma=3b^2(1+r)H$. Then, the continuity equation
(\ref{consq2}) for the dark energy can be written in the standard
form
\begin{eqnarray}
&&\dot{\rho}_D+3H\rho_D(1+w^{\mathrm{eff}}_D)=0.\label{consqeff}
\end{eqnarray}
Substituting Eq. (\ref{wDInt}) in  Eq. (\ref{wef}), we find
\begin{eqnarray}\label{wDeff}
w^{\mathrm{eff}}_D=-\frac{1}{3}-\frac{2\alpha}{3}-\frac{2\sqrt{\Omega_D}}{3c}\cos
y,
\end{eqnarray}
From Eq. (\ref{wDeff})  we see that with the combination of
Brans-Dicke field and HDE, the effective equation of state,
$w^{\mathrm{eff}}_D$, can cross the phantom divide. For instance,
taking $\Omega_D= 0.73$ for the present time and $c=1$, the lower
bound of Eq. (\ref{wDeff}) is $
w^{\mathrm{eff}}_D=-\frac{2\alpha}{3}-0.9$. Thus for $\alpha
>0.15$ we have $w^{\mathrm{eff}}_D<-1$. Therefore, the Brans-
Dicke field plays a crucial role in determining the behaviour of
the effective equation of state of interacting HDE. It is
important to note that in standard HDE ($\alpha=0$) it is
impossible to have $w^{\mathrm{eff}}_D$ crossing $-1$
\cite{Setare1}. Returning to the general case (\ref{wDInt}), we
see that when the interacting HDE is combined with the Brans-Dicke
scalar field the transition from normal state where $w_D >-1 $ to
the phantom regime where $w_D <-1 $ for the equation of state of
interacting dark energy can be more easily achieved for than when
resort to the Einstein field equations is made. In the absence of
the Brans- Dicke field ($\alpha=0$), Eq. (\ref{wDInt}) restores
its respective expression in non-flat standard cosmology
\cite{wang2}
\begin{eqnarray}
w_D=-\frac{1}{3}-\frac{2\sqrt{\Omega_D}}{3c} \cos y-b^2
{\Omega_D}^{-1}\left(1+\Omega_k-\Omega_{BM}\right)\label{wDIntstand}.
\end{eqnarray}
Next, we examine the deceleration parameter, $q=-\ddot{a}/(aH^2)$.
Substituting $w_D$ from Eq. (\ref{wDInt}) in Eq. (\ref{q1}), one
can easily show
\begin{eqnarray}
q&=&\frac{1}{2\alpha+2}\left[(2\alpha+1)^2+2\alpha(\alpha\omega-1)+\Omega_k-(2\alpha+1)\Omega_D-\frac{2}{c}{\Omega^{3/2}_D}\cos
y \right. \nonumber\
\\
&&
\left.-3b^2\left(1+\Omega_k-\Omega_{BM}+2\alpha\left(1-\frac{\alpha\omega}{3}\right)\right)\right]\label{q2Int}.
\end{eqnarray}
If we take  $\Omega_D= 0.73$ and $\Omega_k\approx 0.01$ for the
present time and $c=1$, $\alpha\approx1/ \omega$, $\omega=10^4$,
$\cos y\simeq 1$, $\Omega_{BM}\approx0.04$ and  $b=0.1$, we obtain
$q=-0.5$ which is again compatible with recent observational data
\cite{Daly}. When $\alpha=0$, Eq. (\ref{q2Int}) reduces to the
deceleration parameter of the interacting HDE in Einstein gravity
\cite{wang2}
\begin{eqnarray}
q&=&\frac{1}{2}(1+\Omega_k)-\frac{\Omega_D}{2}-\frac{{\Omega^{3/2}_D}}{c}\cos
y-\frac{3b^2}{2}\left(1+\Omega_k-\Omega_{BM}\right)\label{q3Int}.
\end{eqnarray}
We can also obtain the evolution behavior of the dark energy.
Taking the derivative of Eq. (\ref{OmegaD}) and using Eq.
(\ref{Ldot}) and relation ${\dot{\Omega}_D}=H{\Omega'_D}$, we find
\begin{eqnarray}\label{OmegaD2}
{\Omega'_D}=2\Omega_D\left(-\frac{\dot{H}}{H^2}-1+\frac{\sqrt{\Omega_D}}{c}\cos
y \right),
\end{eqnarray}
where the dot is the derivative with respect to time and the prime
denotes the derivative with respect to $x=\ln{a}$. Using relation
$q=-1-\frac{\dot{H}}{H^2}$, we have
\begin{eqnarray}\label{OmegaD3}
{\Omega'_D}=2\Omega_D\left(q+\frac{\sqrt{\Omega_D}}{c}\cos y
\right),
\end{eqnarray}
where $q$ is given by Eq. (\ref{q2Int}). This equation describes
the evolution behavior of the interacting HDE in Brans-Dicke
cosmology framework. In the limit of standard cosmology
($\alpha=0$), Eq. (\ref{OmegaD3}) reduces to its respective
expression in HDE model \cite{wang2}
\begin{eqnarray}\label{OmegaD4}
{\Omega'_D}=\Omega_D\left[(1-\Omega_D)\left(1+\frac{2\sqrt{\Omega_D}}{c}\cos
y\right) -3b^2(1+\Omega_k-\Omega_{BM})+\Omega_k\right].
\end{eqnarray}
For flat universe, $\Omega_k=0$, and Eq. (\ref{OmegaD4}) recovers
exactly the result of \cite{wang1}.
\section{Summary and discussion\label{CONC}}
In summary, we studied the interacting holographic model of dark
energy  in the framework of Brans-Dicke cosmology where the HDE
density $\rho_{D}= {3c^2}/(8\pi G L^{2})$ is replaced with
$\rho_{D}= {3c^2\phi^2 }/({4\omega L^2})$. Here
$\phi^2={\omega}/({2\pi G_{\mathrm{eff}}})$, where
$G_{\mathrm{eff}}$ is the time variable Newtonian constant. In the
limit of Einstein gravity, $G_{\mathrm{eff}}\rightarrow G$. With
this replacement in Brans-Dicke theory, we found that the
accelerated expansion will be more easily achieved for than when
the standard HDE is considered.  We obtained the equation of state
and the deceleration parameter of the HDE in a non-flat universe
enclosed by the event horizon measured on the sphere of the
horizon with radius $L=ar(t)$. Interestingly enough, we found that
the combination of Brans-Dicke and HDE can accommodate $w_D = -1 $
crossing for the equation of state of \textit{noninteracting} dark
energy. For instance, taking $\Omega_D= 0.73$ for the present time
and $c=1$, the lower bound of $w_D$ becomes $
w_D=-\frac{2\alpha}{3}-0.9$. Thus for $\alpha>0.15$ we have
$w_D<-1$. This is in contrast to Einstein gravity where the
equation of state of \textit{noninteracting} HDE cannot cross the
phantom divide $w_D=-1$ \cite{Li}. When the interaction between
dark energy and dark matter is taken into account, the transition
from normal state where $w_D >-1 $ to the phantom regime where
$w_D <-1 $ for the equation of state of HDE can be more easily
accounted for than when resort to the Einstein field equations is
made.

In Brans-Dicke theory of interacting HDE, the properties of HDE is
determined by the parameters $c$, $b$ and $\alpha$ together. These
parameters would be obtained by confronting with cosmic
observational data. In this work we just restricted our numerical
fitting to limited observational data. Giving the wide range of
cosmological data available, in the future we expect to further
constrain our model parameter space and test the viability of our
model. The issue is now under investigation and will be addressed
elsewhere.

\acknowledgments{I thank the anonymous referee for constructive
comments. This work has been supported by Research Institute for
Astronomy and Astrophysics of Maragha, Iran.}


\begin{thebibliography}{99}

\bibitem{Rie} A. G. Riess, et al., Astron. J.  116, 1009 (1998);\\
  S. Perlmutter, et al.,   Astrophys. J.  517, 565 (1999);\\
  S. Perlmutter, et al.,   Astrophys. J.  {598}, 102 (2003); \\
 P. de Bernardis, et al.,   Nature  {404}, 955 (2000).


\bibitem{Pad} T. Padmanabhan, Phys. Rep.  380 (2003) 235;\\
P. J. E. Peebles,  B. Ratra,  Rev. Mod. Phys. 75 (2003) 559;\\
E.J. Copeland, M. Sami, S. Tsujikawa, Int. J. Mod. Phys. D 15
(2006) 1753.

\bibitem{Suss1}  G. 't Hooft, gr-qc/9310026;\\ L. Susskind, J. Math. Phys. 36 (1995)
6377.

\bibitem{Coh}  A. Cohen, D. Kaplan, A. Nelson, Phys. Rev. Lett. 82 (1999)
4971.



\bibitem{Li} M. Li, Phys. Lett. B 603 (2004) 1.
\bibitem{Suss2} W. Fischler, L. Susskind,
hep-th/9806039.
\bibitem{Huang} Q. G. Huang, M. Li, JCAP 0408 (2004) 013.


\bibitem{Hsu} S. D. H. Hsu, Phys. Lett. B 594 (2004) 13.



\bibitem{HDE} E. Elizalde, S. Nojiri, S.D.
Odintsov, P. Wang, Phys. Rev. D 71 (2005) 103504;\\  B. Guberina,
R. Horvat, H. Stefancic, JCAP 0505 (2005) 001;\\ B. Guberina, R.
Horvat, H. Nikolic, Phys. Lett. B 636 (2006) 80;\\ H. Li, Z. K.
Guo, Y. Z. Zhang, Int. J. Mod. Phys. D 15 (2006) 869;
\\ Q. G. Huang, Y. Gong, JCAP 0408 (2004) 006;\\
J. P. B. Almeida, J. G. Pereira, Phys. Lett. B 636 (2006) 75;
\\  Y. Gong, Phys. Rev. D 70 (2004) 064029; \\ B. Wang, E.
Abdalla, R. K. Su, Phys. Lett. B 611 (2005) 21.


\bibitem{Setare} M. R. Setare, S. Shafei, JCAP 09 (2006) 011;\\  M. R. Setare, E. C. Vagenas, Phys. Lett. B 666 (2008) 111;\\
H. M. Sadjadi, arXiv:0902.2462;\\ M. R. Setare, E. N. Saridakis,
Phys. Lett. B 671 (2009) 331;\\ M. Jamil, E. N. Saridakis, M. R.
Setare, arXiv: 0906.2847.

\bibitem{Seta1}
M. R. Setare, Eur. Phys. J. C 50 (2007) 991;\\ M. R. Setare, JCAP
0701 (2007) 023;\\ M. R. Setare, Phys. Lett. B 654 (2007) 1;\\
M. R. Setare, Phys. Lett. B 642  (2006) 421.

\bibitem{Setare1} M. R. Setare, Phys. Lett. B 642 (2006)
1.

\bibitem{Xin} X. Zhang, F. Q. Wu,  Phys. Rev. D 72 (2005)
043524;\\ X. Zhang, F. Q.  Wu, Phys. Rev. D 76 (2007) 023502;\\
Q. G. Huang, Y. G. Gong, JCAP 0408 (2004) 006;\\ K. Enqvist, S.
Hannestad, M. S. Sloth, JCAP 0502 (2005) 004;\\ J.y. Shen, B.
Wang, E. Abdalla, R.K. Su, Phys. Lett. B 609 (2005) 200.

\bibitem{Feng} B. Feng, X. Wang, X. Zhang, Phys. Lett. B 607 (2005) 35;\\
H.C. Kao, W.L. Lee, F.L. Lin, Phys. Rev. D 71 (2005) 123518;\\ J.
Y. Shen, B. Wang, E. Abdalla, R. K. Su, Phys. Lett. B 609 (2005)
200.


\bibitem{Huang1} Q.G. Huang, M. Li, JCAP 0503 (2005) 001.

\bibitem{Pav1} D. Pavon, W. Zimdahl, Phys. Lett. B 628 (2005) 206.

\bibitem{Fara} V. Faraoni, \textit{Cosmology in Scalar-Tensor Gravity}, Kluwer,
Boston, (2004); \\ E. Elizalde, S. Nojiri, S. D. Odintsov, P.
Wang, Phys. Rev. D 71 (2005) 103504; \\ S. Nojiri, S. D. Odintsov,
Gen. Relativ. Gravit. 38 (2006) 1285;\\  R. Gannouji, et al., JCAP
0609 (2006) 016.

\bibitem{BD} C. Brans and R. H. Dicke, Phys. Rev. 124 (1961) 925.

\bibitem{Pavon2} N. Banerjee, D. Pavon, Phys. Lett. B 647 (2007) 447.

\bibitem{Setare2} M. R.  Setare, Phys. Lett. B 644 (2007) 99.


\bibitem{other} H. Kim, H. W. Lee, Y. S. Myung
 Phys. Lett. B 628 (2005) 11;\\ Y. Gong, Phys. Rev. D 70 (2004)
 064029;\\ B. Nayak, L. P. Singh, arXiv:0803.2930;\\ L. Xu, J.
Lu and W. Li, arXiv :0804.2925;\\ L. Xu, J. Lu  and W. Li, arXiv
:0905.4174.


\bibitem{Ame} L. Amendola, Phys. Rev. D 60 (1999)  043501; \\ L. Amendola, Phys. Rev. D 62 (2000) 043511;
 \\ L. Amendola and C. Quercellini, Phys. Rev. D 68
(2003)  023514; \\ L. Amendola and D. Tocchini-Valentini, Phys.
Rev. D 64 (2001)  043509 ;\\ L. Amendola and D. T. Valentini,
Phys. Rev. D 66 (2002)  043528.

\bibitem{Zim} W. Zimdahl and D. Pavon, Phys. Lett. B 521 (2001) 133;\\ W. Zimdahl and D. Pavon, Gen. Rel. Grav. 35
(2003) 413;\\ L. P. Chimento, A. S. Jakubi, D. Pavon and W.
Zimdahl, Phys. Rev. D 67 (2003)  083513.

\bibitem{wang1} B. Wang, Y. Gong and E. Abdalla, Phys. Lett. B 624
(2005) 141.

\bibitem{wang2} B. Wang, C. Y. Lin and E. Abdalla, Phys. Lett. B 637
(2005) 357.



\bibitem{wang3} B. Wang, C. Y. Lin. D. Pavon and E. Abdalla, Phys. Lett. B 662
(2008) 1;\\ W. Zimdahl and D. Pavon, Class. Quantum  Grav. 24
(2007) 5461;\\ D. Pavon and A. A. Sen, arXiv:0811.1446.

\bibitem{Oli} G. Olivares, F. Atrio, D. Pavon, Phys. Rev. D 71 (2005) 063523.



\bibitem{spe} D. N. Spergel, Astrophys. J. Suppl. 148 (2003)
175;\\ C. L. Bennett, et al.,  Astrophys. J. Suppl. 148 (2003)
1;\\
M. Tegmark, et al., Phys. Rev. D 69 (2004) 103501;\\ U. Seljak, A.
Slosar, P. McDonald, JCAP 0610 (2006) 014;\\ D. N. Spergel, et
al., Astrophys. J. Suppl. 170 (2007) 377.

\bibitem{Arik}  M. Arik, M.C. Calik, Mod. Phys. Lett. A  21 (2006)
1241;\\ M. Arik, M. C. Calik  and M. B. Sheftel, gr-qc/0604082.


\bibitem{Will} C. M. Will, \textit{Theory and Experiment in Gravitational
Physics},\\ Cambridge University Press, Cambridge, (1993).

\bibitem{Bert} B. Bertotti, L. Iess and P. Tortora, Nature, 425 (2003)
374.

\bibitem{Aca} V. Acquaviva, L. Verde, JCAP 12 (2007) 001.

\bibitem{c} Since $\omega\geq10^4$ we find that $\alpha\approx
1/\omega=10^{-4}$, thus Eq. (26) reduces practically to $
w_D=-\frac{1}{3}(1+2\alpha)-\frac{2\sqrt{\Omega_D}}{3c}\approx
-\frac{1}{3}-\frac{2\sqrt{\Omega_D}}{3c}$ which is exactly the
Li's result. Thus, the Li's argument \cite{Li} in favor of $c=1$
holds here.
\bibitem{Daly} R.A. Daly et al., Astrophysics J. 677 (2008) 1.

\bibitem{Tsu} S. Tsujikawa, M. Sami, Phys. Lett. B 603 (2004) 113.

\bibitem{Dor} M. Doran, J. Jackel, Phys. Rev. D 66 (2002) 043519.

\end{thebibliography}
\end{document}